\documentclass[floatfix,%
 amsmath,amssymb,
 aps,
prb,
twocolumn
]{revtex4-2}

\usepackage{graphicx}
\usepackage{dcolumn}
\usepackage{bm}
\usepackage{xcolor}
\newcommand{\RT}[1]{\textcolor{red}{#1}}

\begin{document}


\title{Dynamic compression of glassy GeO$2$ up to the TPa range and first observation of shock induced crystallization}

\author{R. Torchio$^{1}$,  J.A. Hernandez$^{1}$, A. Cordone$^{2}$, S.Balugani$^{1}$, T. Vinci$^3$, C. Pepin$^4$, N. Sevelin-Radiguet$^1$, E. Guillam$^3$, F. Dorchies$^{5}$, A.Benuzzi$^{3}$}
\affiliation{$^1$European Synchrotron Radiation Facility, 6 Rue Jules Horowitz, BP220, 38043 Grenoble Cedex, France}
\affiliation{$^2$ Politecnico di Milano}
\affiliation{$^3$ Laboratoire pour l'Utilisation des Lasers Intenses, Ecole Polytechnique, CNRS, Commissariat à l'Energie Atomique, Sorbonne Université, 91128 Palaiseau, France.} 
\affiliation{$^4$CEA, DAM, DIF, 91297 Arpajon Cedex, France}
\affiliation{$^5$Université Bordeaux, CNRS, CEA, CELIA, UMR 5107, Talence 33400, France}



 Second institution and/or address\\
 This line break forced
%



\begin{abstract}
In this work we present an extensive study of glassy GeO$_2$ under laser induced dynamic compression. New  VISAR and SOP data provide the extension of Hugoniot EoS up to the TPa range for this material including temperature measurements. Reflectivity data at both 532 and 1064 nm wavelenght are also reported. In the low compression range we observe changes of the optical properties from transparent, to opaque, to metallic state. 
The second part of this work describes a further laser shock experiment combined with in-situ X-ray diffraction. Here we observe, for the first time, the laser shock-induced crystallization of glassy GeO$_2$ to a structure compatible with the rutile phase at pressure higher than 20 GPa and melting occurring at around 75 GPa.
\end{abstract}

\maketitle


\section{Introduction}

Germanium dioxide (GeO$_2$) is an important material for technology and material science as well as for geophysics and planetary science. It can exist in three polymorphic phases at ambient conditions: quartz, rutile and glass. Thanks to their mechanical and optical properties, GeO$_2$ polymorphs can be applied in a number of devices from solar cells, to encapsulating materials for Ge semiconductors, to technical glasses undergoing extreme conditions of pressure and temperature.

Generally, GeO$_2$ is regarded as a chemical and structural analog of SiO$_2$, in particular, the structure of glassy GeO$_2$ is a good model for amorphous or molten SiO$_2$ 
. 

As SiO$_2$ analogues, the interest in investigating GeO$_2$ polymorphs under extreme conditions of pressure and temperature extends to Earth's and planetary science. SiO$_2$ is the major constituent of the continental crust, the main component in silicate melts present in Earth’s mantle and also expected to be present in deep Earth and to be a key constituent of recently discovered rocky extrasolar planets 
, where thermodynamic conditions reach up to multi Mbar pressures and several thousand Kelvin temperatures, in the so called Warm Dense Matter range. 

The density increase mechanism upon atomic scale compaction for the amorphous and crystalline phases of SiO$_2$ and GeO$_2$ has important geophysical and geochemical implications for planetary evolution models, in particular regarding the stability and mobility of deep silicate melts in both Earth and rocky planets.

Recently an increase in coordination from 6 to 7 for glassy GeO$_2$ - i.e. beyond the coordination of the crystalline rutile counterpart -  was reported in the 40-90 GPa range using static compression 
and molecular dynamic simulations 
, as well as in the molten GeO$_2$ quartz phase upon laser-shock compression in the 50-100 GPa range 
.


Under static compression up to 80 GPa and below 2000K, the GeO$_2$ phase diagram exhibits a smaller number of polymorphs and all GeO$_2$ phases have a larger sensitivity to pressure, undergoing pressure-induced changes at  lower pressures than their SiO$_2$ analogues 
. Investigating GeO$_2$ as a model for SiO$_2$  thus present several advantages, given the simpler and more accessible phase diagram, and the higher atomic number that allows for hard X-ray based characterization.



The GeO$_2$ phase diagram is still largely unexplored especially at very extreme conditions of pressure and temperature, i.e towards the Warm Dense Matter regime. GeO$_2$ polymorphs have been investigated up to 130 GPa in cold static compression and up to 70GPa/1600K and 45GPa/3200K using static compression in the laser-heated diamond anvil cell (LH-DAC) mostly coupled to synchrotron XRD 
Flying plates dynamic compression experiments reach 160 GPa and lack any direct microscopic probe 
. Recently, two experiments have been performed on quartz GeO$_2$ using laser-shock compression coupled to time-resolved XRD up to 104 GPa at MEC-LCLS 
and laser-ramp compression coupled to time resolved-XRD up to 884 GPa 
. Sholmerich reports the sequence quartz-rutile-melting in the 19-104 GPa range. The laser ramp experiment reports HP-PdF2-type phase in the 150-440 GPa range and emergence of a new phase likely cotunnite-type above. 
The rutile and glassy GeO$_2$ phases have not been investigated in the multiMbar range.

In this work we present an extensive study of glassy GeO$_2$ under laser induced dynamic compression. 

Since few decades, a lot of effort is being devoted to the combination of laser-driven dynamic compression experiments to brilliant X-rays probes, both at large laser facilities 
- where X-rays are produced  in situ - and at brilliant X-rays facilities 
- where more compact lasers are coupled to the X-ray beamlines - with the aim to unveal the microscopic properties of the matter under very extreme conditions. In this context, the existence of reliable Equations of State (EoS) - a thermodynamic equation relating state variables, such as pressure, volume, temperature or internal energy -  becomes crucial.  In fact, the complexity of such experiments translates into strong constraints to the target design, so that in most of the cases, only one shock velocity can be measured and a known EoS is needed to locate the microscopic findings in the P/T diagram.

In the first part of this work, we present VISAR and SOP data providing the extension of Hugoniot EoS up to the TPa range for this material including temperature measurements. Reflectivity data at both 532 and 1064 nm wavelenght are also reported. In the low compression range we observe changes of the optical properties from transparent, to opaque, to metallic state.

The second part of this work describes a further laser shock experiment combined with in-situ X-ray diffraction. Here we observe, for the first time, the laser shock-induced crystallization of glassy GeO$_2$ to a structure compatible with the rutile phase at pressure higher than 20 GPa and melting occurring at around 75 GPa. Our data are then discussed and compared with available results from static compression and dynamic compression for the other GeO$_2$ polymorphs and SiO$_2$ glassy phase.

\section{METHODS}
This section is dedicated to the description of experimental setups and diagnostics, target designs and methods employed to analyse the data.  The scientific results are then presented and discussed in Section \ref{Results}.

\subsection{Experimental setup}
\begin{figure}
\includegraphics [width=8.5 cm]{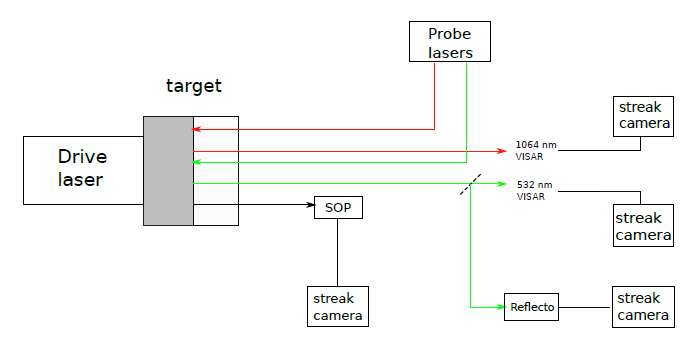}
\caption{Scheme of the experimental setup for the EoS measurements.}
\label{fig:setupEoS}
\end{figure}

\begin{figure}
\includegraphics [width=8.5 cm]{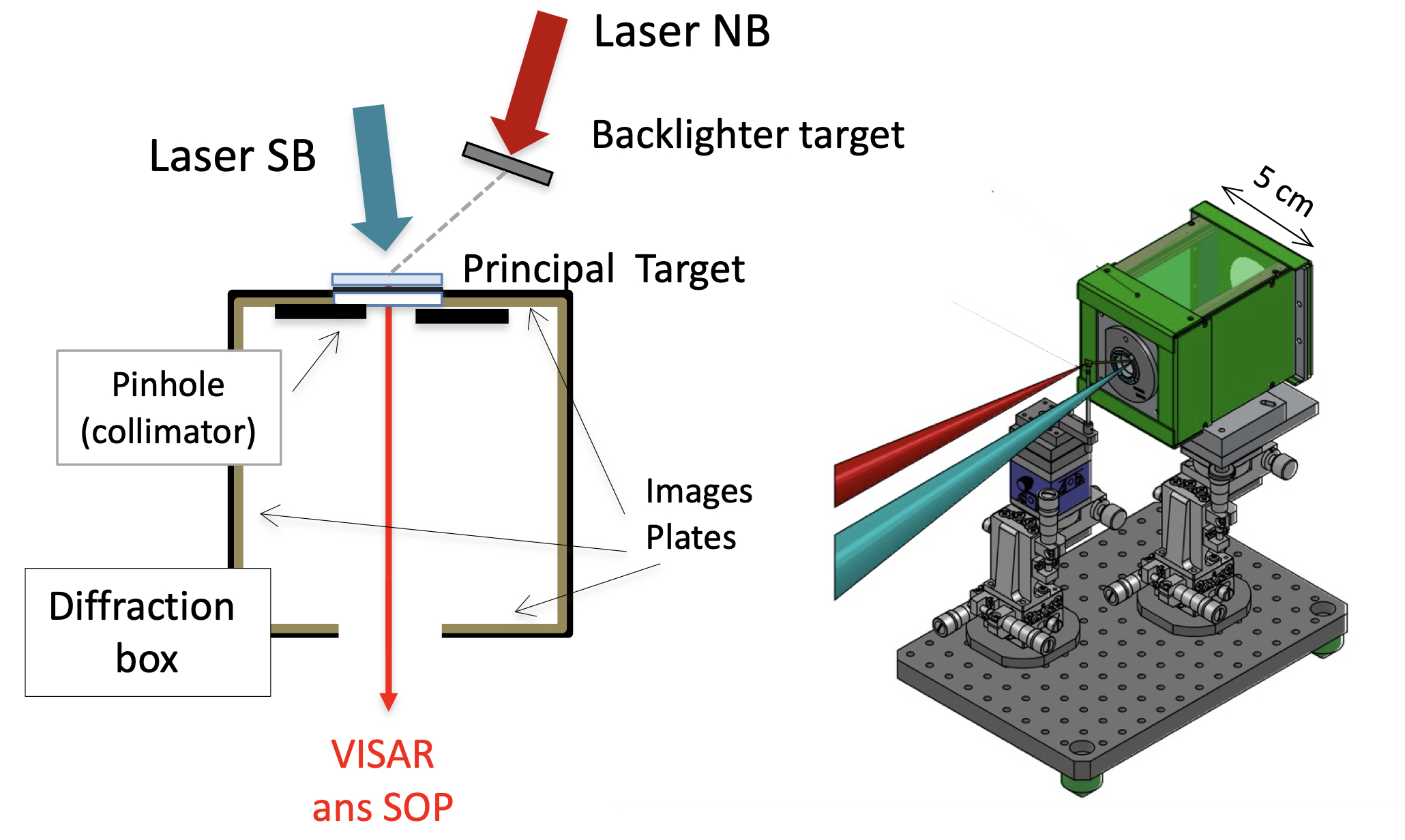}
\caption{Scheme of the experimental setup for the X-ray diffraction campaign.The collimation of the X-ray source was ensured by a 500 $\mu$m tantalum or lead pinhole.}
\label{fig:setup}
\end{figure}
Experiments were performed at the LULI2000 laser facility at the LULI Laboratory, where we dispose of two long pulse beams, South Beam (SB) and North Beam (NB). The set-ups of EoS and X-ray diffraction campaigns are shown schematically in Figure \ref{fig:setupEoS} and \ref{fig:setup}  respectively. For the EoS campaign, the frequency-doubled drive laser beams (wavelength = 0.532 $\mu$m)  had a duration of 2 or 5 ns and both were spatially smoothed with Hybrid Phase Plates (HPP) providing laser beam size on target of 500 or 800 $\mu$m allowing to reach laser intensities between 2 10$^{12}$ W/cm$^2$ to 3 10$^{14}$ W/cm$^2$ and so to collect data in a large domain of pressure conditions. Here, the main diagnostics were VISAR (velocity interferometer system for any reflector) at two wavelengths (532 and 1064nm) and a streaked optical pyrometer (SOP). Moreover, one arm of the VISAR at 532 nm is deviated before the interferometer to collect reflectivity measurements on a dedicated streak camera called "Reflecto".

For the X-ray diffraction campaign, one beam (SB) was used to compress the target. The SB duration was 10 ns and the focal 500 $\mu$m (always spatially smoothed by HPP), giving a laser intensity up to  10$^{13}$ W/cm$^2$.  Here, the goal was to study structural changes in a low pressure regime (up to the melting). The second beam (NB) was used with a duration of 1 ns and focalised on a iron or copper target to generate the X-ray source (the He$\alpha$ line at 6.7 keV and 8.4 keV respectively). In this case, the main diagnostic was the diffraction box (see section G). On the rear side, VISAR and SOP diagnostics were installed to have an independent measurement on the shock parameters and to check on each shot the probe time versus the shock dynamics.  

\begin{figure}
\includegraphics [width=8.5 cm]{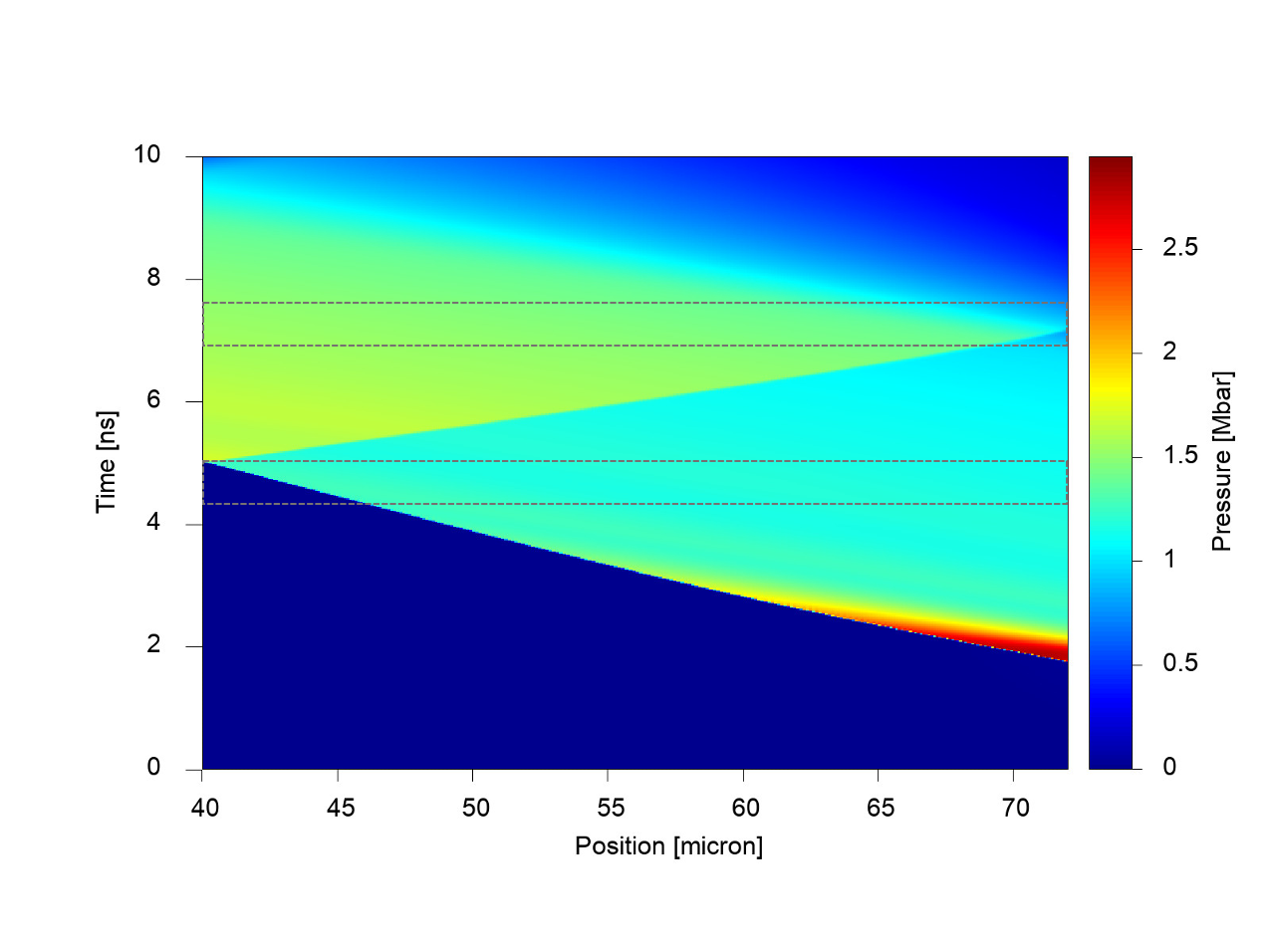}
\caption{Example of hydrodynamic simulation at 2 10$^{12}$ W/cm$^2$ laser intensity. The dotted areas represent the probed times with X-rays on the Hugoniot and in re-shock conditions. The X-ray pulse duration is around 0.8 ns.}
\label{fig:sim}
\end{figure}

The principle of the X-ray diffraction diagnostic is the same described by Rygg et al. Here, the box was made of aluminium structure and the dimension was 5x5x7 cm$^3$. A lead screen (2 mm) was positioned on the front side to shield the coronal emission and reduce the noise. The FujiTM BAS-MS image plates (IP) covered each side of the box included the rear side where a hole was done for SOP and VISAR diagnostics. The distance between the backlighter(BL) and the main target(MT) was 2 cm and the angle between the axis BL/MT and the shock direction was 40°. The MT was mounted to a 150 $\mu$m thick tantalum or lead plate with a 500 $\mu$m diameter collimating aperture. In addition to providing x-ray collimation, diffraction from the edge of this aperture provides reference calibration lines on the IPs corresponding to ambient uncompressed Ta or lead. The thickness of GeO$_2$ in MT was chosen to optimize the diffraction signal and to probe uniform conditions. The transmitted diffracted signal is maximized when the thickness of the target is around the optical depth at the energy of the X ray source. In our case, the optical depth is 34$\mu$m and 61$\mu$m at 6.7 and 8.4 keV. The thickness was fixed at 32$\mu$m (corresponding to a  thickness crossed by X-rays of 41.7 $\mu$m, considering the incidence angle) to ensure uniform conditions and also due to thickness limitation in the coating capability. Indeed to design the target, we checked steadiness of the shock and then uniform conditions, by performing some hydrodynamical simulations. An example is shown in Fig. \ref{fig:sim}. Using 1 ns pulse to generate the source, we estimated a duration of 0.7-0.8 ns of the X-ray probe. The dotted areas in the figure are the regions probed in this experiment, corresponding to conditions along the Hugoniot or along re-shocked Hugoniot induced by diamond layer on the rear side.  This layer also provided a further calibration of the X-ray diffraction. On each shot, the X-ray source was monitored by two Von Hamos spectrometers. In figure \ref{fig:difffr} left panel we present a typical image obtained on rear side IP of the target at ambient conditions. Since GeO$_2$ is in vitreous state at standard conditions, we detect only diffraction lines from Ta and diamond.

\subsection{Target design}
Specific targets were designed for the EoS experiment and combined laser-shock XRD experiment.
The EoS target for the high energy (HE) regime, is made of a GeO$_2$ deposition (30 $\mu$m) performed by DEPHIS company on a SiO$_2$ quartz commercial substrate (50 $\mu$m). A thin layer of CH and a thick layer of Al are glued on the front side to act as ablator (CH) and shield to x-rays radiation (Al) from the coronal plasma following laser ablation.
This design is used in the range where both SiO$_2$ and GeO$_2$ are loaded to a metallic state. 

\begin{figure}
\includegraphics [width=8 cm]{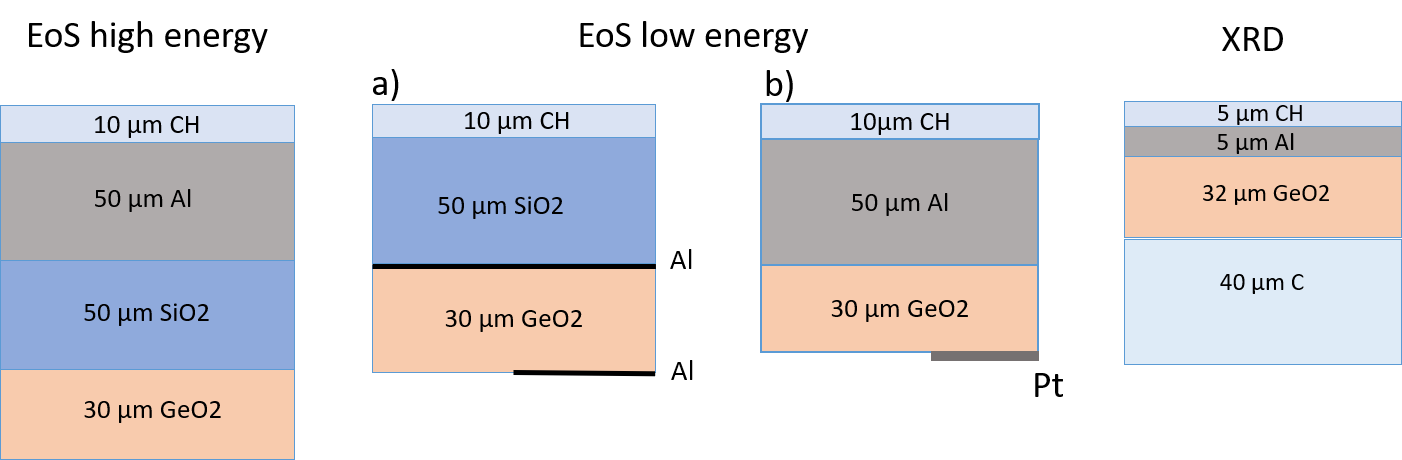}
\caption{targets designs for EoS (left and center) and combined shock/XRD experiment (right).}
\label{targets}
\end{figure}
Two different designs were used for the low energy regime.
Target a) (see Fig. \ref{targets})  was made of a  30 $\mu$m GeO$_2$ deposition on a 50 $\mu$m SiO$_2$ quartz substrate coated with a flash of Al (200 nm). Half of the rear face of GeO$_2$ is also coated with 200 nm Al. On the front side, 10 $\mu$m of CH are glued on the SiO$_2$ layer for the ablation. Target b) is similar but GeO$_2$ was deposited on a 50 $\mu$m Al substrate and part of the rear face of GeO$_2$ was coated with 200 nm Pt.
These target were used in the low energy regime to observe the optical properties changes of GeO$_2$ from transparent to opaque to metallic as the shock loading is increased, and for absolute EoS measurements. 

Finally the target optimised for laser shock combined to in-situ XRD in transmission geometry was made of 32 $\mu$m deposition of GeO$_2$ on a diamond window of 40 $\mu$m while thin layers of CH/Al were respectively deposited and glued on the ablation side to serve as ablator (CH) and ease the detection of the shock entrance in GeO$_2$ (Al).

\subsection{Hugoniot measurements}
When a material is submitted to a shock wave, the density, pressure, and internal energy density of the shocked state ($\rho_1$, P$_1$,
and e$_1$, respectively) are linked to those of the initial state ($\rho_0$, P$_0$, and e$_0$) by the Rankine-Hugoniot relations
\begin{equation}\label{eq:rho_rankine}
    \rho_1 = \rho_0\frac{U_s}{U_s-U_p}
\end{equation}
\begin{equation}\label{eq:rankine_pressure}
    P_1 = P_0 + \rho_0 U_sU_p
\end{equation}
\begin{equation}\label{eq:rankine_energy}
    e_1-e_0 = \frac{1}{2}(P_0+P_1)\Big(\frac{1}{\rho_0} - \frac{1}{\rho_1}\Big)
\end{equation}

where U$_s$ is the shock front velocity and U$_p$ is the material or particle velocity.

If a material in an initial state 0 is compressed through a single shock loading, the pressure reached will be
uniquely determined by the the initial state and the density reached after compression, i.e. the final state will be found on the Hugoniot curve H($\rho_1$, P; $\rho_0$, P$_0$) =0.

Therefore, in order to obtain Hugoniot  data of a shocked material, knowing its initial state, it is thus necessary to measure two variables out of 5 of the Rankine-Hugoniot relations directly or through a reference sample as described in the following sections.

\subsubsection{Hugoniot measurements in the Low Energy regime}

\begin{figure}
\includegraphics [width=8.5 cm]{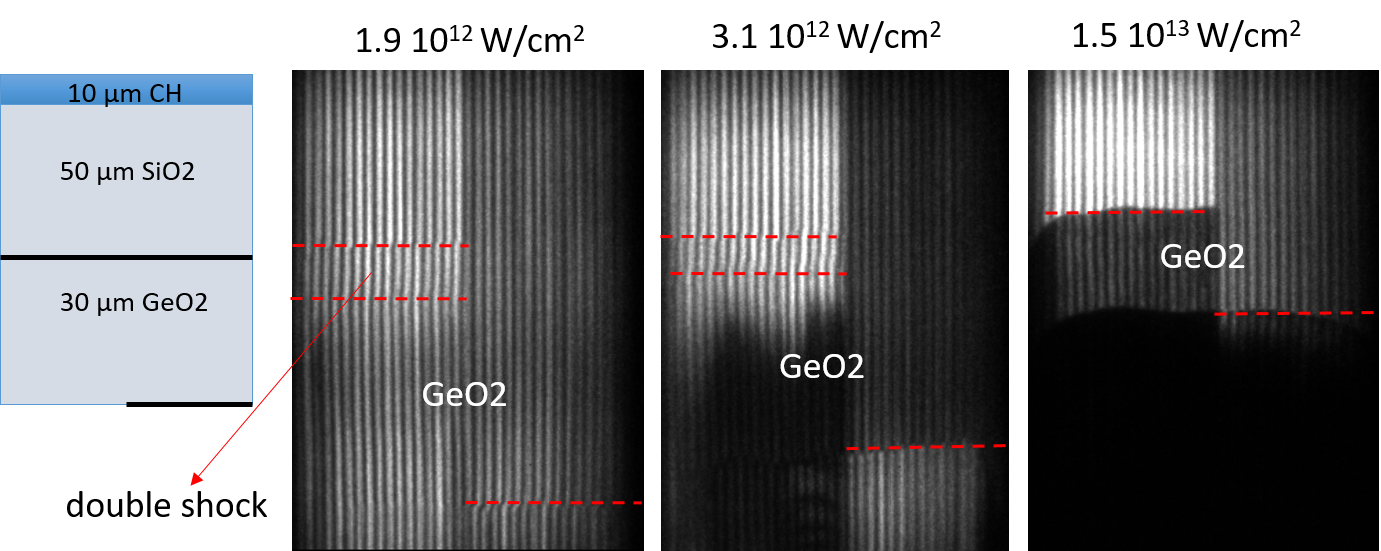}
\caption{Compression induced changes of optical properties for the LE target: from transparent (left) to opaque (center) to metallic (right). The two dashed lines on the left side of the target indicate the double shock structure due to the elastic wave originated in the SiO$_2$ quartz substrate. The dashed line on the right side indicates the free surface.}
\label{fig:LE}
\end{figure}

In the low energy regime we observed the optical properties changes of glassy GeO$_2$ from transparent to opaque to metallic, as the laser intensity is increased (see Fig. \ref{fig:LE}).
In the case of target a), in the very low compression range, within 50 GPa, a double shock structure in observed in GeO$_2$ (see Fig. \ref{fig:LE}, left and central panel). This is due to the elastic precursor wave of the quartz SiO$_2$ substrate, previously reported in the 10-60 GPa range 
. 
The first shock wave that enters the $GeO_2$ is the elastic one  with an almost constant velocity. The second shock is the plastic wave which is slower and its velocity  depends on the intensity of the drive laser. In fact, as the drive laser intensity  is increased the elastic shock enters the $GeO_2$  at the same instant while the second shock, the plastic one, arrives earlier (left and central panel of Fig. \ref{fig:LE}). The two shock waves merge inside the GeO$_2$ layer and no double shock structure is observed at the free surface.   The elastic limit of $GeO_2$ has been reported at quite low pressure (about 4 GPa by I. Jackson \textit {et al.} 
), and was not observed in the investigated range.

At compressions where the two shocks are merged, it is possible to obtain absolute Hugoniot data. From the left side of the target, the shock velocity U$_s$ of GeO$_2$ can be estimated as average transit velocity (right of Fig. \ref{fig:LE}), knowing the GeO$_2$ layer thickness and measuring the transit time from the VISAR images.  From the right side of the target, where a flash of Al has been deposited, the particle velocity U$_p$ can be obtained from the measurement of the free surface velocity in the free surface approximation: $U_p \approx \frac{U_{free}}{2}$.
This approximation 
, only holds in the low compression range below  metallization/melting, where the entropy jump across the shock front is negligible.

To avoid the issue of the elastic precursor of SiO$_2$, target b) was also used for the low compression regime. In the range where GeO$_2$ becomes opaque, EoS data were obtained in the same way as for target a). 

The shots that left the GeO$_2$ in a transparent state needed a more careful analysis as here the VISAR laser travels in a transparent, but partially shocked medium-requiring the knowledge of the shocked index of refraction. Here an apparent velocity, $U_{app}$, is measured from the VISAR data. 
Assuming a linear dependency of the index of refraction on the density, $n(\rho)=a(\rho-\rho_0)+n_0$, one can prove that:
\begin{equation}
    U_p=U_{app}/(n_0-a\rho_0)
\end{equation}
 where the factor $a$ can be derived by linear interpolation of refractive index data as a function of density from 
 .


\subsubsection{Hugoniot measurements in the High Energy regime}
11 shots have been performed in the high energy configuration with laser intensities ranging from 2 10$^{13}$ W/cm$^2$ to 2.8 10$^{14}$ W/cm$^2$ using phase plates of 800 $\mu$m (up to 10$^{14}$) and  500 $\mu$m (above).

\begin{figure}
\includegraphics [width=8 cm]{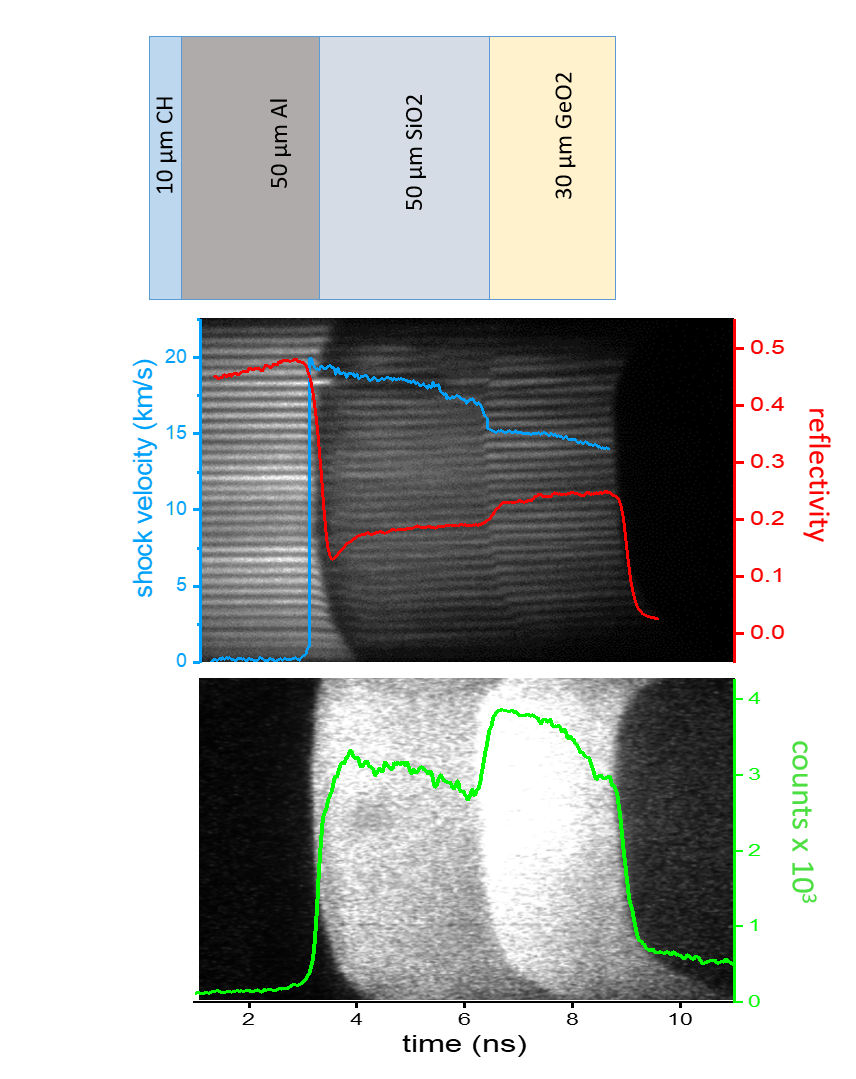}
\caption{Scheme of the HE target (top). Center: example of VISAR image with extracted shock velocity (clear blue) U$_s$ and reflectivity (red) curves as the shock wave goes through the SiO$_2$ and GeO$_2$ layers. Bottom example of SOP image with extracted self emission intensity curve (green).}
\label{HE}
\end{figure}

At sufficiently high laser intensities, the shock wave generated by the laser ablation of the CH layer, can produce a reflecting shock front in initially transparent materials such as SiO$_2$ and GeO$_2$. In this case, the VISAR diagnostics allow the measurements of the shock velocity U$_s$ in both SiO$_2$ and GeO$_2$. Interferometry images were recorded by the two VISARs working at 532 and 1064 nm wavelenght. Fig.\ref{HE} shows an example of VISAR image obtained for the HE target at 10$^{14}$ W/cm$^2$ laser intensity. Fringes jumps are observed as the shock front crosses the Al/SiO$_2$ and the SiO$_2$/GeO$_2$ interface and the corresponding velocities are calculated (clear blue line), taking into account the material ambient refractive index.

\begin{figure}
\includegraphics [width=8 cm]{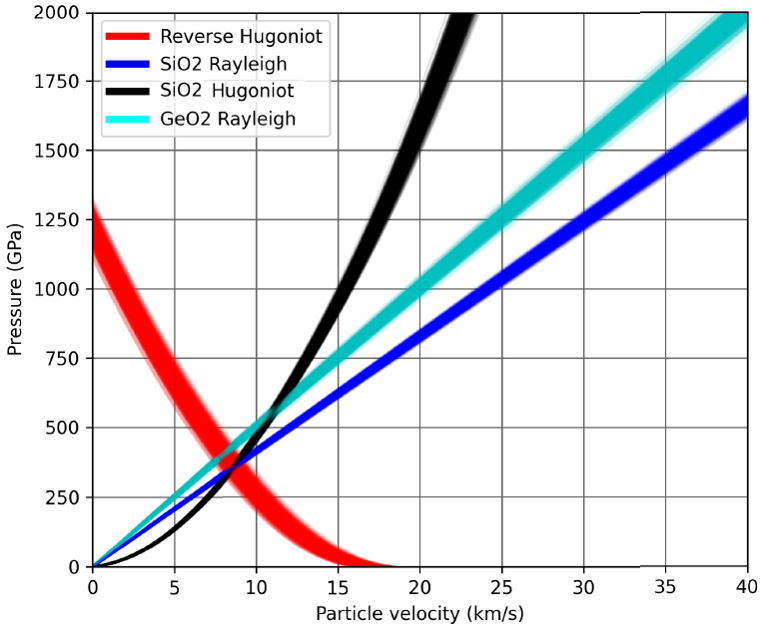}
\caption{Impedance matching graph: the shock Hugoniot of SiO$_2$ quartz
is in black, the reverse Hugoniot in red, and Rayleigh lines $\rho_0U_sU_p$ for GeO$_2$ and SiO$_2$ are in clear blue and blue respectively.}
\label{mismatch}
\end{figure}

Since the Hugoniot EoS of SiO$_2$ quartz is well characterized in the literature 
, it is possible to obtain the  particle velocity (U$_p$) and pressure (P) of GeO$_2$ using the impedance mismatching technique 
. The underlying principle is dynamic equilibrium at the SiO$_2$/GeO$_2$ interface. In fact, when the shock reaches the GeO$_2$ layer, as the impedance of the
GeO$_2$ is greater than the impedance of SiO$_2$, it loads up GeO$_2$ to a higher pressure state. A re-shock wave then is generated propagating backwards into the SiO$_2$ to ensure continuity of pressure and particle velocity across the interface. As the two U$_s$ are measured and the SiO$_2$ Hugoniot is known, a graphical construction allows to derive P and U$_p$ in GeO$_2$, as illustrated in Fig. \ref{mismatch}.

The incident shock state (U$_p$, P) is given by the intersection of the Hugoniot curve of SiO$_2$ with its Rayleigh line $\rho_0U_sU_p$ while the transmitted shock state is given by the intersection of the reverse Hugoniot with the Rayleigh of GeO$_2$. The SiO$_2$ quartz Hugoniot has been taken from Brigoo et al. 
. All the parameters errors as well as the GeO$_2$ ambient density and refractive index error have been progagated using a Montecarlo method that translates into a band of lines as in Fig. \ref{mismatch}.

The full set of visar data, from LE to HE regime is reported in  Fig.\ref{fig:Hugoniot} in comparison to previous literature data 
, in the Us/Up and $\rho$/P planes.

\subsection{Reflectivity and temperature}

Reflectivity (R) and temperature (T) data were acquired in the range where GeO$_2$ becomes reflecting. Reflectivity values are deduced from the signal intensity changes in the VISARs and Reflecto images as the shock crosses the targets. In the case of a reflecting shock, the reflectivity (at
the wavelength of the probe laser) is due to a mismatch of complex refractive index between the shock compressed material (index n) and the pre-compressed material (index n0).
$$R=|n-n_0|/|n+n_0|$$
The quartz SiO$_2$ layer is used as an in situ calibrant. Indeed, as the Us-R relations for both 532 and 1064 nm wavelength are known 
, the measured R/U$_s$ values for SiO$_2$ can be anchored, to obtain reflectivity values for GeO$_2$.

Temperature measurements were performed using Streaked Optical Pyrometry (SOP) on an independent optical path. The target thermal emission is recorded as a function of time and related to the temperature through the Planck law in the grey body approximation:
\begin{equation}
    L(\lambda,T) = \epsilon(\lambda,T) A \frac{1}{exp(hc/k_b\lambda T)-1}
    \label{planck}
\end{equation}
where  L($\lambda$,T) is the spectral radiance of a grey body at temperature T and wavelength $\lambda$ and A is a calibration factor that takes into account the optical transfer function of the optical system and the detector response.
Targets emissivities $\epsilon(\lambda,T)$ at 532 nm are obtained from the reflectivities values thought the  Kirchhoff’slaw.
\begin{equation}
    \epsilon(\lambda,T) = 1 - R(\lambda,T)
\end{equation}

and assumed to be constant over the probe laser and SOP $\lambda$ working range. Here again SiO$_2$ is used as an in situ calibrant: the temperature calibration parameter is found by achoring the measured emissivity and U$_s$ data in SiO$_2$ to reference literature data 
. 

Reflectivity and temperature data are reported in Figs \ref{fig:reflectivity} and \ref{fig:temp}.

\subsection{In situ-X-ray diffraction} 
\begin{figure}
\includegraphics [width=8 cm]{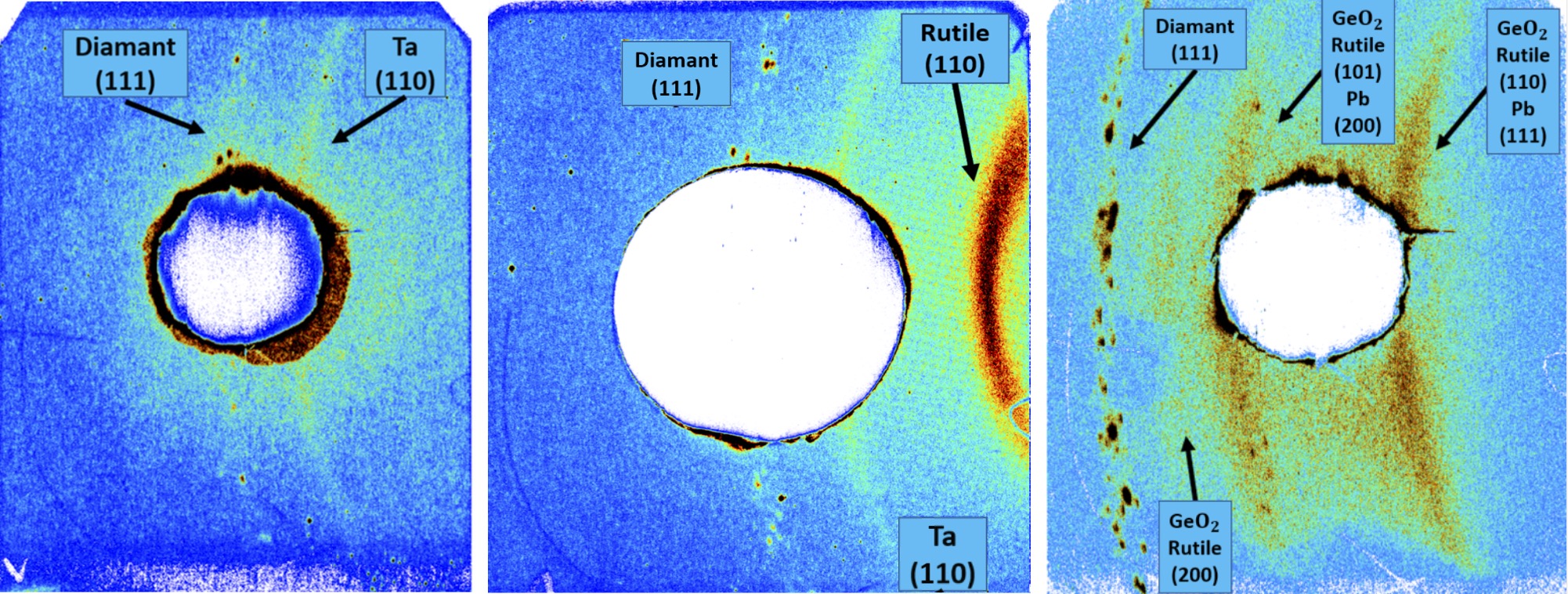}
\caption{X-ray diffraction data collected on rear side box IP. Left: ambient conditions; center and right: under shock conditions with X-ray source at 8.4 and 6.7 keV respectively.}
\label{fig:difffr}
\end{figure}

\begin{figure}
\includegraphics [width=8 cm]{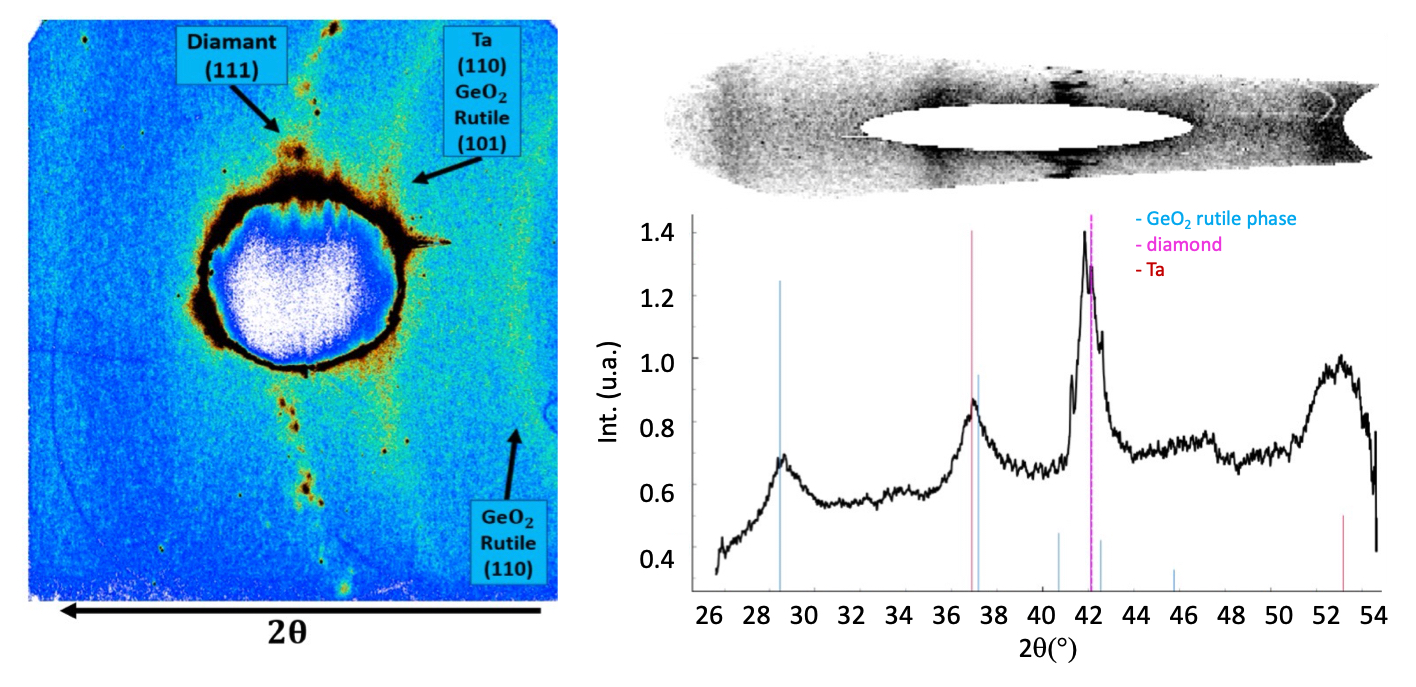}
\caption{Left: raw XRD image; right: correspondent XRD pattern obtained with Dioptas. \RT{can we put a bigger label in the right panel?}}
\label{fig:spectreDioptas}
\end{figure}

In transmission geometry, the X-ray beam diffraction occurs in a cone whose aperture angle 2$\theta$ is related to the intereticular distance of the crystal planes (d-spacing) and to the wavelength of the X-ray beam by the Bragg's law. The diffraction cone is then recorded on the Imaging Plates (IP), protected by a plastic foil, lining the inside of a diffraction box. The lines on the IPs correspond to the intersection between the different planes of the IPs and the diffraction cone. The diffraction cone can be  reconstructed over a wide range of 2$\theta$ (25° and 110° approximately). In order to properly identify the crystalline lines in the shocked material, calibrants are used as diffraction references. The calibrants used here are the CVD diamond layer of the main target and tantalum or lead pinholes glued on the rear side target, as shown in figure \ref{fig:setup}. The CVD diamond lines are easily identified by the fact that the lines are composed of diffraction dots corresponding to the polycrystalline structure of the diamond. As example, we show  X-ray diffraction data corresponding to IP on the rear side of the box (VISAR side) on figure \ref{fig:difffr}. We present here on the left the data of a cold target, and on the center and on the right data corresponding to a shocked target at similar compressions, with two different X-ray sources, respectively at 8.4 keV and 6.7 keV.  At ambient conditions (left) no diffraction lines from the GeO$_2$ are observed due to its vitreous phase. Under shock conditions (center and right panels), the clear appearance of new lines can be observed. 
The XRD images calibration, taking into account the diffraction box geometry, and data processing has been performed using the Dioptas code.
An example is shown in \ref{fig:spectreDioptas}.

\begin{figure}
\includegraphics [width=8.5 cm]{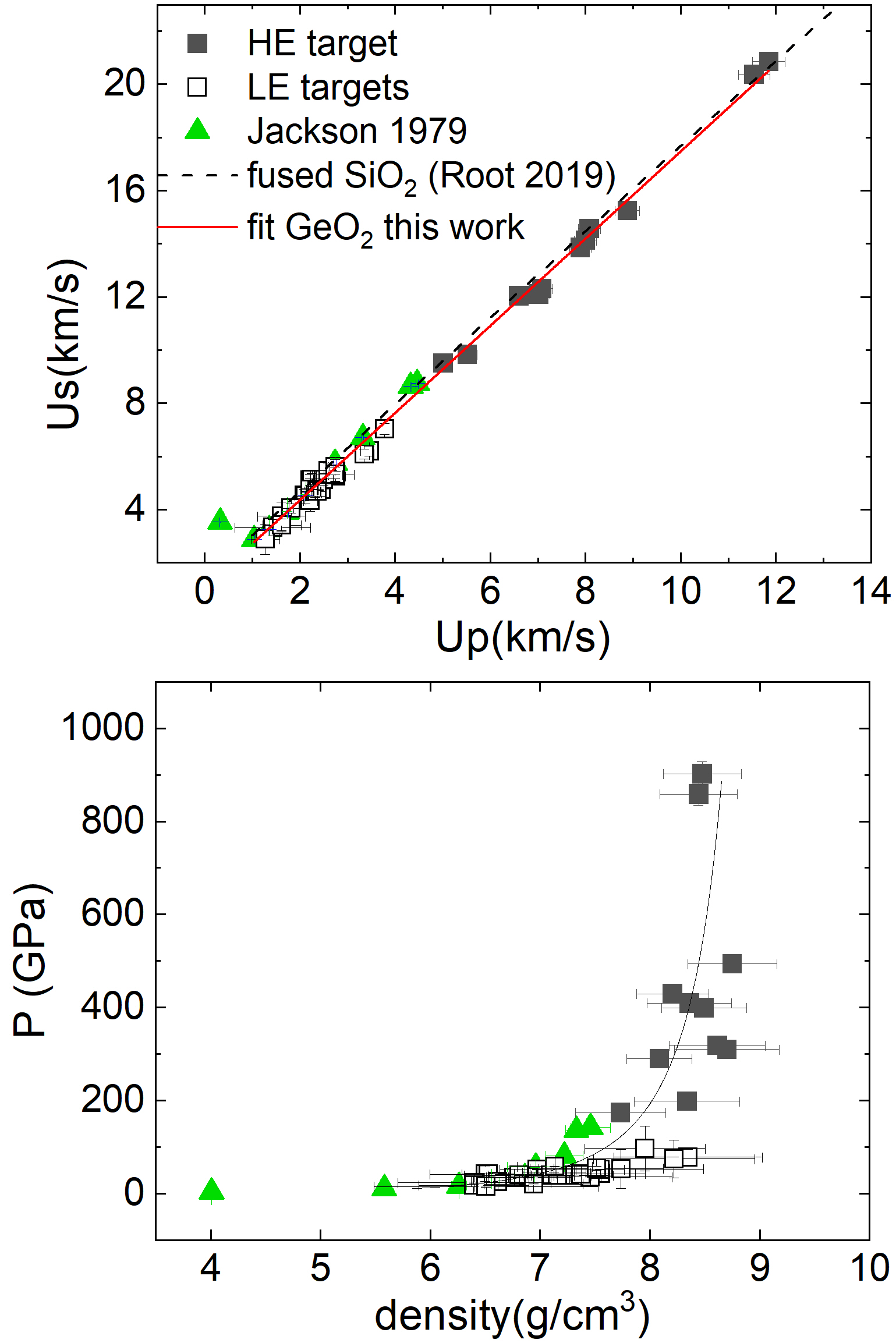}
\caption{Hugoniot data in the Us/Up (top panel) and $\rho$/P planes (bottom panel). Black full and open squares  are measurements from the visar with higher sensitivity (2$\omega$) in the HE and LE configuration respectively; green triangle are from 
}
\label{fig:Hugoniot}
\end{figure}

\section{Results and discussion}
\label{Results}
In this paper we report Hugoniot data for glassy GeO$_2$ up to the TPa range and microstructural characterization by in-situ XRD up to around 100 GPa.

\subsection{Hugoniot curve, reflectivity and temperature}


Our obtained Hugoniot data are reported in Fig. \ref{fig:Hugoniot} together with previous data from 
, where the flying plate technique was used. In that work, a discontinuity was observed in the Hugoniot data, mostly visible in the $\rho$/P plane, and attributed to transition to a denser phase.  No clear discontinuity is observable in our Hugoniot data, likely due to the larger error in the LE range. On the other hand our in-situ XRD data clearly confirm the crystallization to a different phase as will be discussed below.\\

A linear fit of our full set of Us/Up data, 
merged to those of 
gives:
$$Us=1.07(\pm{0.03})+1.64(\pm{0.01})Up$$

here, only the first point of the Jackson series, corresponding to GeO$_2$ elastic limit, has been excluded.
The fit is reported with the data in the upper panel of Fig. \ref{fig:Hugoniot} together with the one recently reported for fused SiO$_2$ (Root2019). A fit to a cubic polynomial, as done for fused SiO$_2$ (Root 2019), does not significantly improve the fit quality. 
The Us/Up trend for for fused SiO$_2$ and GeO$_2$ results to be very close, within the error bars.
 \\

\begin{figure}
\includegraphics [width=8.5 cm]{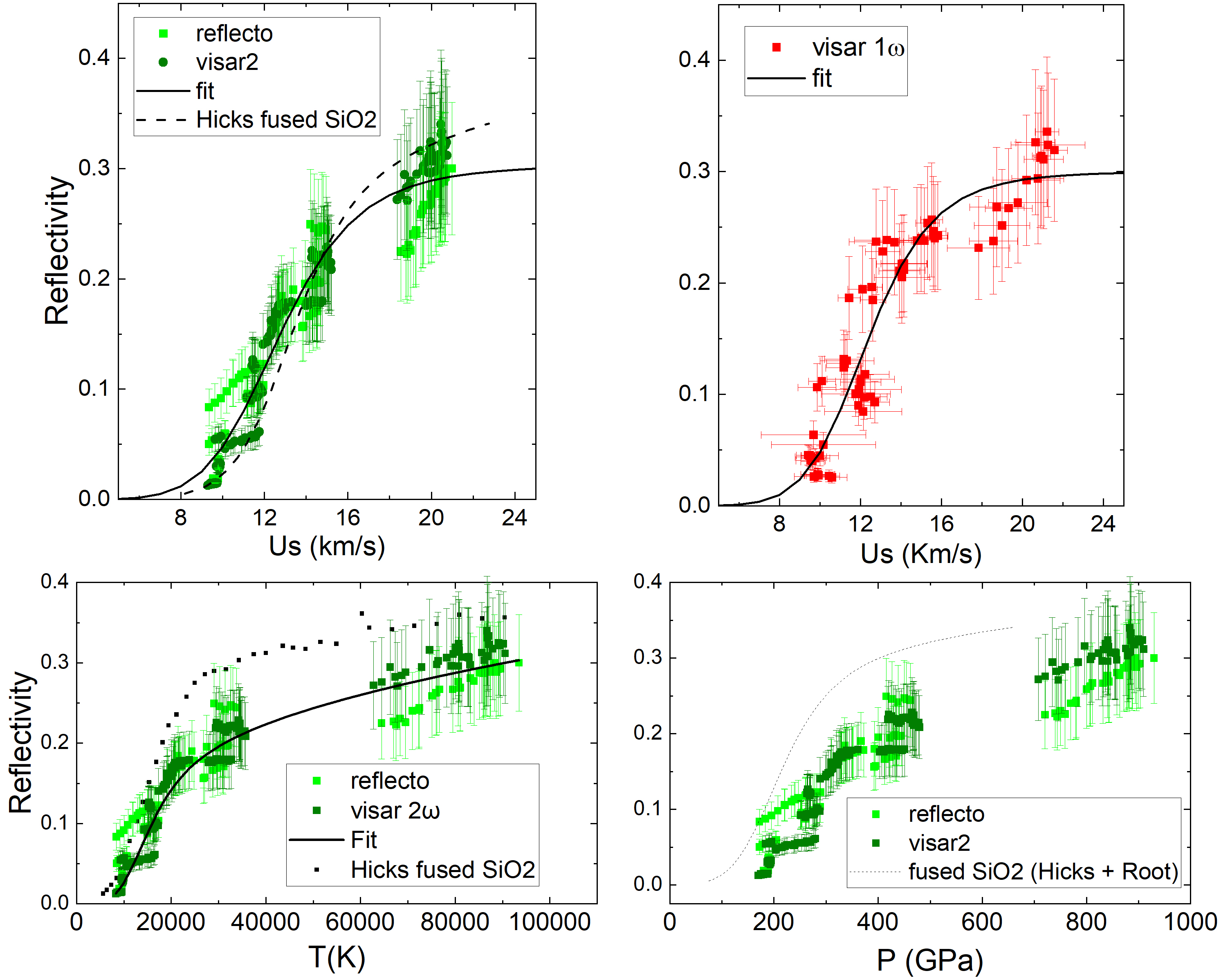}
\caption{Top panels: reflectivity data from the visar images at 2$\omega$ and the non interferometric arm (reflecto) (top left panel, dark and clear green scatters) and from visar images at 1$\omega$ (top right panel, red scatters), as a function of U$_s$. Bottom panels: reflectivity data at (2$\omega$) as a function of temperature (left) and pressure (right). All data at 2$\omega$ are compared to literature data from 
}
\label{fig:reflectivity}
\end{figure}

Reflectivity data at 532 (2$\omega$) and 1064 nm (1$\omega$) are shown in Fig.\ref{fig:reflectivity}.
\begin{figure}
\includegraphics [width=8.5 cm]{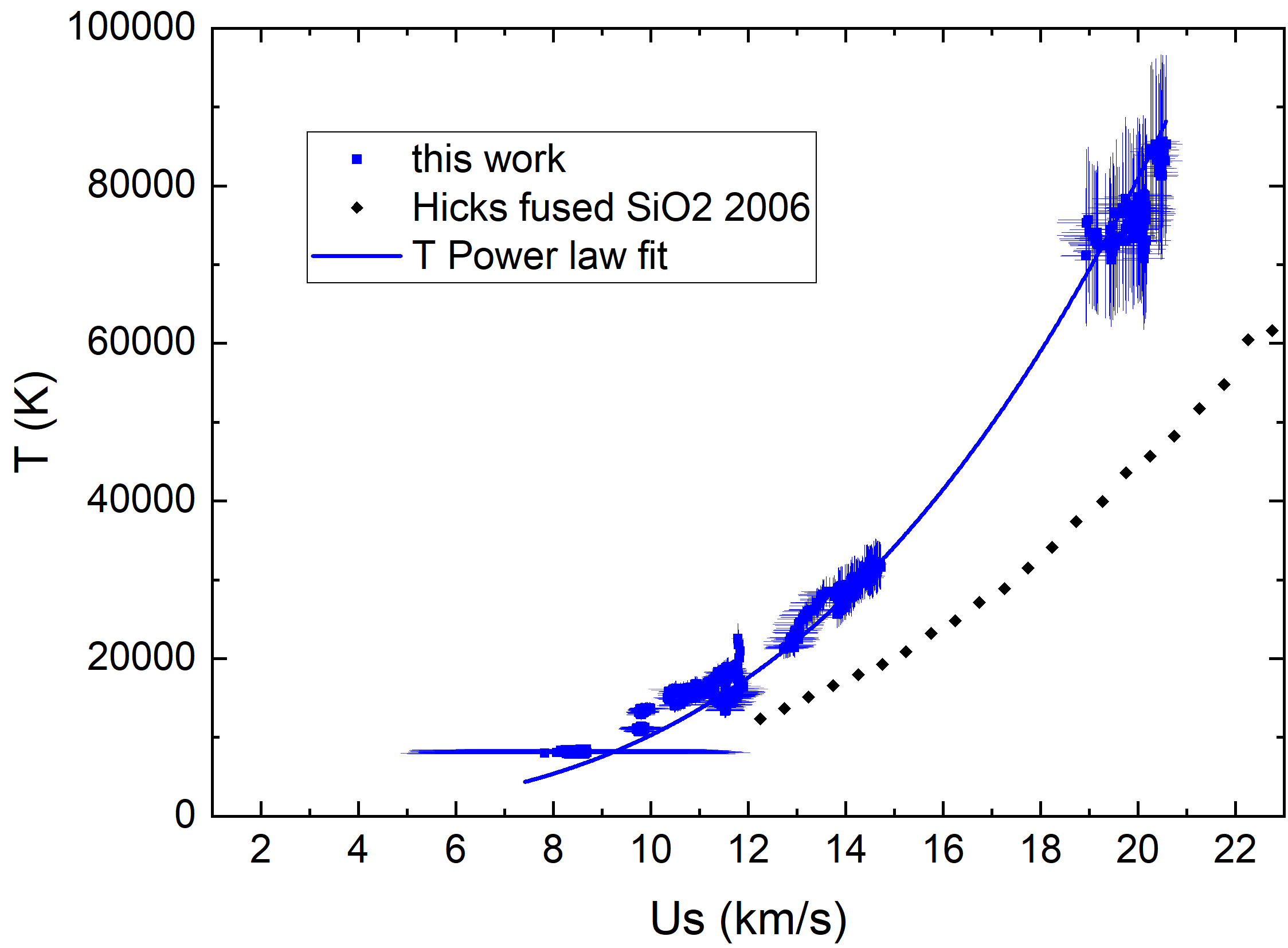}
\caption{Temperature data in the Us and P plane.}
\label{fig:temp}
\end{figure}

Fitting to a Hill function:
\begin{equation}
    R = a + (b - a)\frac{U_s^{c}}{U_s^{c} + d^{c}}
    \label{GeO2-1064}
\end{equation}

gives the parameters reported in table \ref{refl}.

\begin{table}[b]
\caption{Hill function parameters for the reflectivity data}
\label{refl}
\begin{ruledtabular}
\begin{tabular}{ccccc}
 & a & b & c & d\\
\hline\hline
532 nm & 0(0.11) & 0.303(0.005) & 6.8(0.6) & 12.8(0.1)\\
1064 nm& 0(0.04) & 0.30(0.015) & 6.7(1.8) & 12.4(0.5)\\
\end{tabular}
\end{ruledtabular}
\end{table}

For the 2$\omega$ data, we compare to data of fused SiO$_2$ from 
reported in 
, in the R(U$_s$), R(T) and P(T) planes. The two materials reflectivity depict quite similar behavior as a function of U$_s$ within the error bar. 
In the R/P plane the reflectivity onset for glass GeO$_2$ is at higher pressures, as expected for a denser material. In the R/T plane, the reflectivity onset is again quite similar within the error bar, however higher values are reached in fused SiO$_2$.

A selection of temperature data along the Hugoniot is reported in Fig.\ref{fig:temp} as a function of U$_s$. Our data are well reproduced by a power law:
$$T(K)=300+18.4(\pm{0.3})U_{s}^{2.784(\pm{0.006}))}$$
The glass GeO$_2$ Hugoniot temperatures vs U$_s$ result higher with respect to those reported by 
for fused SiO$_2$ in the 12-23 range of U$_{s}$.


\subsection{Microstructural analysis}

\begin{figure}
\includegraphics [width=8.5 cm]{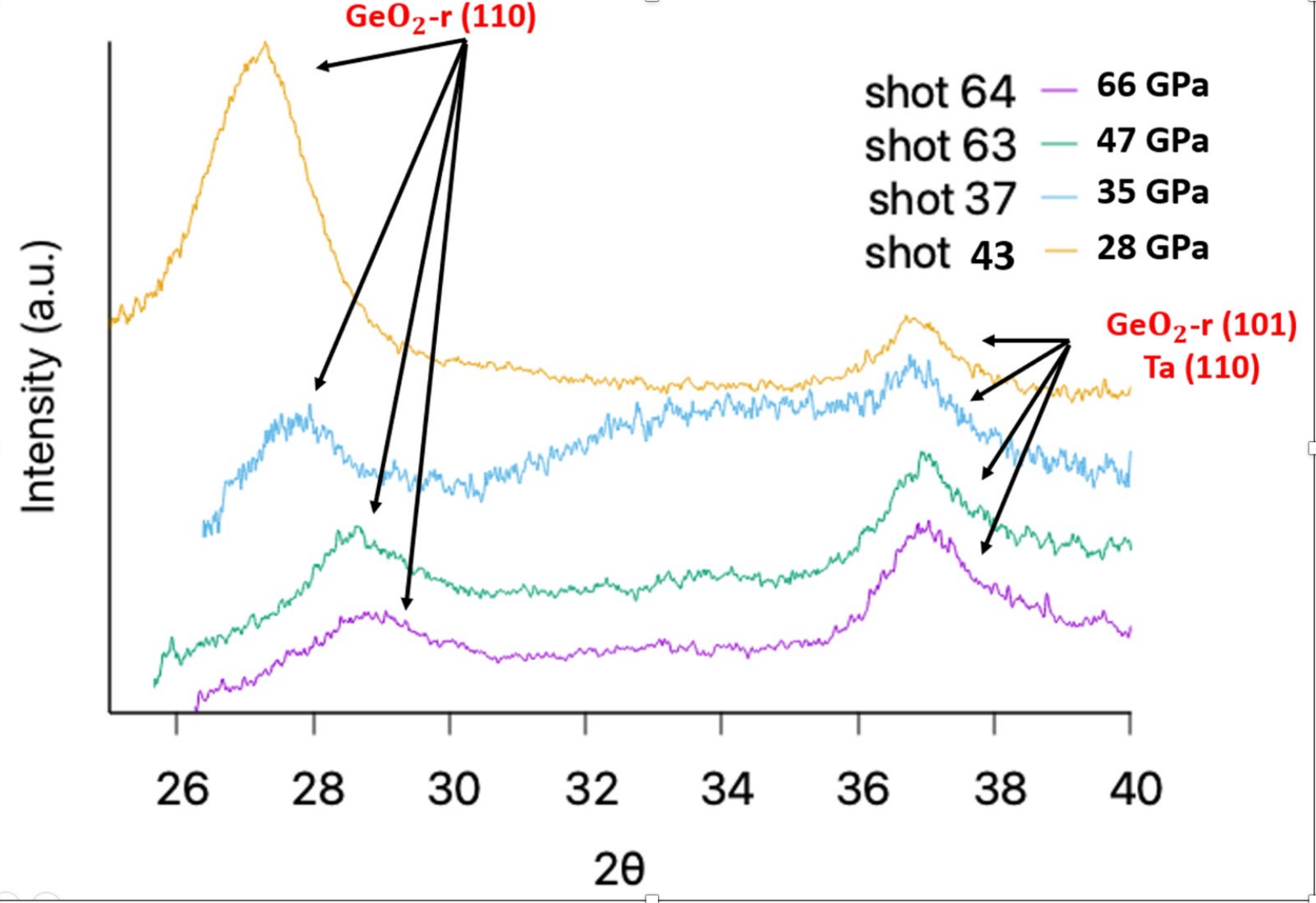}
\caption{XRD patern evolution as a function of pressure.}
\label{fig:shift}
\end{figure}

In-situ XRD diffraction measurements, performed in our second experiment, clearly show the emergence of new Bragg reflections from density around 6.45 g/cm$^3$, corresponding to pressures of around 22 GPa ($\pm$5 GPa), compatible with the crystalline rutile structure (Fig.\ref{fig:spectreDioptas}).
In Fig.\ref{fig:shift}, we show the X-ray diffraction signals as a function of 2$\theta$ for shots corresponding to a shock pressure of GeO$_2$ between 28 and 66 GPa \RT{\textit{why don't we show also lower P data? there should be at 22-23 GPa}}. These pressures were determined by average shock velocity measurements from the VISAR images combined with our Hugoniot data described in the previous session. From the shift of the peaks, we were also able to deduce the GeO$_2$ density variations and check the consistency with the VISAR data. It should be noted that with the shift of the 110 reflection is evident, the observation of shift of the 101 reflection is hindered by the overlapping with a reflection from the Ta pinhole. 

This finding finally confirms shock induced crystallization of glass GeO$_2$  to rutile-like structure, previously predicted from flying plates measurements, where a significant densification had been observed along the shock Hugoniot 
around 20-25 GPa. Other papers suggest the onset of the glass-rutile transition already from 6-8 GPa and a mixed phase region up to 20-25 GPa 
, similarly as observed in static compression 
. Our two data points below 22 GPa do not shown any Bragg reflection for GeO$_2$, however, within the error bar, they are very close in pressure. No data could be acquired at more moderate conditions, therefore, we cannot fully rule out an earlier transition onset.
The observed shock-induced phase transition in glass GeO$_2$ is similar to that of laser shocked fused SiO$_2$ 
and to that of crystalline quartz GeO$_2$ 
where crystallization to rutile phase  was observed at 18-19 GPa upon shock. 

The range of the structural transition to the crystalline phase roughly corresponds to the change in optical properties from transparent to opaque, as the nucleation of nanograins of rutile results in increased light scattering.

The Bragg reflections of the rutile-like phase disappear at densities higher than 7.5 gr/cm$^2$ (P$>$80 GPa) indicating the occurrence of melting or amorphization.  

No literature data are available on melting of glassy GeO$_2$. The melting line of rutile GeO$_2$ is reported only up to 2 GPa by 
.
The onset of shock melting in crystalline quartz GeO$_2$ has been recently reported, by in-situ XRD at LCLS 
, to be at lower pressure, i.e. P$<=$52 GPa. A similar onset pressure was inferred from projectile impacts  and laser-shock measurements for fused SiO$_2$, with a fluid-solid co-existence range up to 70 GPa,  and believed to be delayed by metastable superheating of the shock-induced crystalline phase (Lyzenga, 
).
By comparison with these similar systems and prediction, we infer that the observed loss of crystalline order at  85$\pm$5 GPa corresponds to melting.

\begin{figure}
\includegraphics [width=8.5 cm]{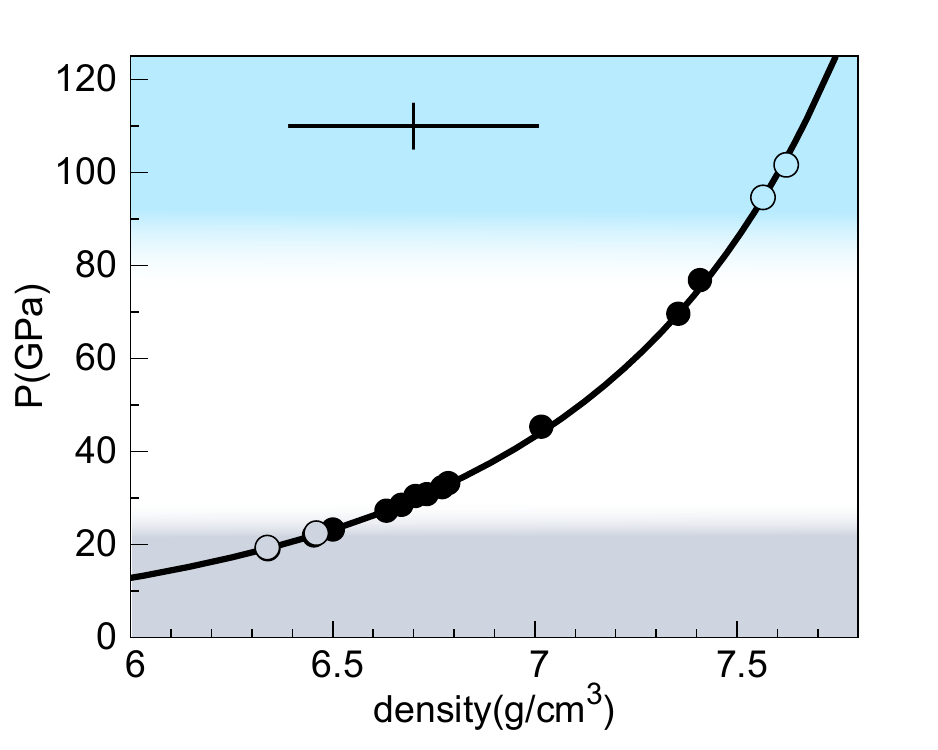}
\caption{Crystallization and melting/amorphization along the principal Hugoniot. Gray points for glass GeO$_2$, black for rutile crystal phase and light blue for molten GeO$_2$. The error bar is displayed on a side for clarity.}
\label{fig:crystal}
\end{figure}

We report crystallization and melting data along the Hugoniot curve in the $\rho$/P plane in Fig.\ref{fig:crystal}.

Interestingly, the loss of crystalline order in GeO$_2$,  anticipate the metallization onset above 100 GPa, as extrapolated by our reflectivity measurements in Fig. \ref{fig:reflectivity}, that would imply, in the case of melting, the presence of poorly electrically conducting liquids close to the melting lines, as previously observed in MgO-SiO2 compounds 
.

\begin{figure}
\includegraphics [width=8.6 cm]{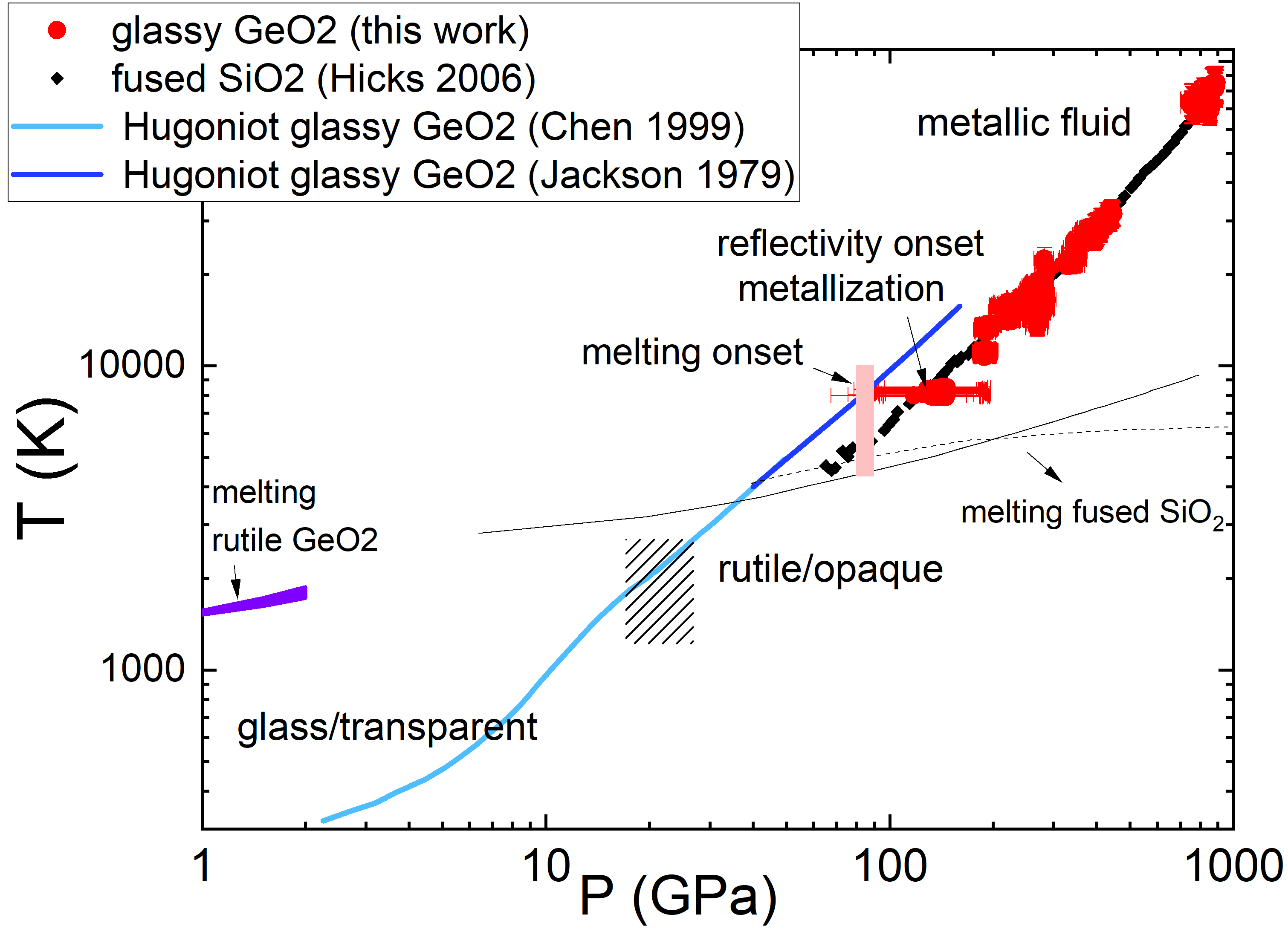}
\caption{Phase diagram of glass GeO$_2$ from this work and previous literature. Red rounds are  new Hugoniot data from this work in the liquid-metallic state. 
Black diamonds are fused SiO$_2$ data from 
.  The melting lines for fused SiO$_2$ are from 
. The melting line for rutile GeO$_2$ is from 
. Clear and navy blue lines are calculated Hugoniot for glass GeO$_2$ from 
. The dash and pink rectangles represent estimated phase boundaries for the glass/rutile and solid/liquid transitions respectively. }
\label{fig:phase diagram}
\end{figure}

Finally we draw a P/T phase diagram for glass GeO$_2$ combining our new results with literature data, in Fig. \ref{fig:phase diagram}.
In the P/T plane, in the liquid range, Hugoniot data for glass GeO$_2$ are found to be almost superimposed to those of fused SiO$_2$ 
, with the exception of very extreme data at 750-900 GPa where shock temperatures for GeO$_2$ results to be higher but still close to those of SiO$_2$ within the error bar (\textit{to add a zoomed figure in SI }). 

Our data are shifted towards higher pressures with respect to calculated Hugoniot from previous works 
. This is likely due to the existence of a melting plateau. Therefore, the data reported from our XRD experiment in the solid phase, showing the glass/rutile/liquid transitions, might be placed on these curves. However, since
 the temperatures could not be measured directly in our XRD experiment, we have indicated rough phase boundaries vs temperature, encompassing both the extrapolation of our P,T data to lower conditions and the curves of 
 . 
The loss of crystalline order, that we interpret as melting, was detected around 85 GPa,  a value quite close to the one observed in fused SiO$_2$ 
. No literature melting data are available for glassy GeO$_2$. The only available data are for the rutile GeO$_2$ (violet curve in the figure), whose extrapolation would cross the Hugoniot curve at quite lower pressure. 
.


\section{Conclusions}
In this paper we provide an exaustive investigation on glassy GeO$_2$, a critical material for technology and planetary science, submitted to laser-induced dynamic compression up to the TPa range.
New Hugoniot data together with independent reflectivity and temperature data are provided and compared to fused SiO$_2$, showing that the two materials share several common properties at extreme dynamic P,T conditions, but also some differences.  Moreover, we provide functional relations that could allow to use glassy GeO$_2$ as an impedance standard.

Our XRD data provide the first direct experimental evidence of shock-induced crystallization to the rutile structure, similarly as observed in static compression, while the material transforms from transparent to opaque. No other phase transitions are observed. Bragg reflections from the rutile-like phase are observed in the 23-80 GPa range, but an earlier onset, as observed in static compression, cannot be fully excluded. The loss of crystalline order, that we interpret as melting, is observed above 80 GPa- quite close to the reported melting of fused SiO$_2$ and much higher then previously predicted melting for glass GeO$_2$. Interestingly, melting anticipates metallization by several tens of GPa. 

The combination of our Hugoniot, reflectivity, temperature and XRD data allows to largely extend the knowledge of the P,T phase diagram of glassy GeO$_2$.


\end{document}